# Performance and failure modes of AI chatbots on a novel concept inventory on relativity in classical mechanics

Eugenio TUFINO (1), Caterina GIOVANZANA (2), Andrea ZAMBONI (2), Pasquale ONORATO (2) and Stefano OSS (2)

*(1) Department of Physics, Informatics and Mathematics, University of Modena and Reggio Emilia, Via G. Campi, 213/A, Modena, Italy*
*(2) Department of Physics, University of Trento, Via Sommarive 14, 38123, Povo (Trento), Italy*

*E-mail: caterina.giovanzana@unitn.it

**Abstract**
AI chatbots are increasingly used by students as study tools in physics, raising practical questions about their reliability on conceptual tasks. Existing evaluations of large language models (LLMs) on physics concept inventories rely almost exclusively on instruments that have been publicly available for years and likely appear in model training data, making it difficult to disentangle physics competence from familiarity with the test items themselves. We address this issue by evaluating three frontier LLMs (GPT-5.2, Gemini 3 Pro, Gemini 3 Flash) on the Classical Relativity Concept Inventory (CRCI), a recently developed and validated 21-item instrument on Galilean relativity that was not publicly available at the time of testing. Each item was administered 30 times per model, and all 1890 responses were qualitatively coded along three dimensions: visual interpretation, physics reasoning, and coordination. Mean accuracy was 97% for Gemini 3 Flash, 89% for Gemini 3 Pro, and 73% for GPT-5.2, compared to 62% for the student sample (N = 267). However, all three models fail completely on a small number of items. The qualitative analysis shows that these failures stem predominantly from misinterpretations of visual content rather than from deficits in physics knowledge, and that LLM errors differ structurally from those of students: when models err, they converge on a single distractor with high consistency, whereas student errors are more broadly distributed. These findings indicate that chatbot reliability on conceptual physics is item-dependent and unpredictable, with direct implications for how concept inventories are administered.

# 1. Introduction

AI-powered chatbots, such as OpenAI 'ChatGPT and Google 'Gemini, are increasingly used by university students as study tools in physics courses. This raises a practical question for instructors: to what extent can these tools reliably handle the physics we teach?
This question is especially relevant for conceptual reasoning, where students might be tempted to use a chatbot as a study companion or even to answer assessment questions.

A growing body of research has begun to address this issue by testing large language models (LLMs) on physics concept inventories — the standardized multiple-choice tests widely used in physics education research to probe conceptual understanding (1) (2). Early studies showed rapidly improving performance on the Force Concept Inventory (FCI) (1), with GPT-4 achieving over 80% correctness (3). At the same time, Wheeler and Scherr (4) observed that the chatbot's incorrect answers often mirrored common student misconceptions, raising the question of whether these systems genuinely reason about physics or simply reproduce patterns from their training data. More recent work by Polverini and Gregorcic (5) (6) (7) introduced a systematic methodology for evaluating chatbot performance. Their approach involved repeatedly submitting each item of the Brief Electricity and Magnetism Assessment (BEMA) (8) and classifying errors as visual, reasoning, or coordination difficulties. Their findings showed that ChatGPT-4o outperforms average university students, while still struggling with items that require images interpretation. Kortemeyer et al. (9) confirmed this pattern on a large scale, testing GPT-4o on 54 concept inventories across 35 languages.

However, all of these studies share a limitation: the instruments they used — the FCI (1), the Test of Understanding Graphs in Kinematics (TUG-K) (10), the BEMA (8), and others — have been publicly available for years, sometimes decades. Their items and answer keys can be found online, making it plausible that they were included in the training data of the models being tested (9). Therefore, we cannot be sure whether high performance reflects genuine physics competence or a sort of memorization.

In this study, we address this issue by testing three frontier chatbots on a concept inventory they could not have seen during training. In particular, we use the Classical Relativity Concept Inventory (CRCI), a 21-item questionnaire on Galilean relativity and non-inertial reference frames that was recently developed and validated at the University of Trento (11). Its validation paper was published in early 2026 — after the knowledge cutoff of all models we tested — and the instrument has not been made freely available online or indexed by repositories such as PhysPort. The physics covered by the survey — reference frames, the principle of relativity, Galilean velocity composition, the weak equivalence principle — is standard textbook material, but the specific questions, scenarios, and distractors are new. This allows us to observe how chatbots handle familiar physics presented in an unfamiliar format.

We tested GPT-5.2, Gemini 3 Pro, and Gemini 3 Flash — three of the most widely accessible frontier chatbots available to university students at the time of this study. Each item was submitted 30 times per model, following the repeated-trial methodology of Polverini and Gregorcic (6), for a total of 1890 responses. Many survey items feature hand-drawn illustrations of physical scenarios that require spatial interpretation, making the test particularly suited for probing the interplay between visual processing and physics reasoning — an area identified as a persistent weakness of current AI systems (5) (6) (9).
In this paper, we present the results of this evaluation. We report overall and item-level performance for all three models, classify the types of errors that emerge from the chain-of-thought outputs, and compare AI performance and distractor choices to those of the 267 university students who took the same test (11). Our findings are informative both for instructors who want to understand the capabilities and limitations of the chatbots their students use, and for the physics education research community investigating AI performance on conceptual physics tasks.

# 2. Setting and research questions

## 2.1 The instrument: the CRCI

The survey consists in a 21-item multiple-choice designed to assess university students' understanding of core concepts in relative motion (11).
The inventory covers three main conceptual areas: (i) Reference Frames, including the general case of transformations between frames with different accelerations; (ii) the Principle of Relativity and Galilean transformations, including trajectory transformations between inertial frames and Galilean velocity composition; and (iii) the weak Principle of Equivalence, covering horizontal motion in free fall scenarios, simple harmonic motion in free fall, and orbital motion.
The development of the CRCI followed established procedures in the concept inventory literature, including cognitive interviews, expert review, pilot studies, and rigorous statistical validation using classical test theory and the Rasch model. The instrument was administered in several years to 267 first-year physics students at the University of Trento.
A crucial feature for the present study is that the CRCI was not publicly available before or during our data collection. At the time the LLMs were tested, the CRCI had not appeared online in any form — neither the questions nor the answer key. This differs much from the situation for established instruments such as the FCI or BEMA, whose items have been widely reproduced and discussed in publicly accessible sources for years.

## 2.2 The models

We tested three frontier LLMs available in December 2025: GPT-5.2 (OpenAI), Gemini 3 Pro (Google), and Gemini 3 Flash (Google). These models were selected primarily because they are among the most widely accessible AI chatbots available to university students.
For the Google family, we set the temperature T = 0.7, consistent with the settings used in previous studies (6) (9), as both Gemini 3 Pro and Gemini 3 Flash allow temperature control. We included both models in order to compare a larger, more capable model (Gemini 3 Pro) with a lighter and faster one (Gemini 3 Flash). As we report in section 4.1, Gemini 3 Flash slightly outperforms the larger Gemini 3 Pro on most items.
For GPT-5.2, the study uses the baseline configuration reasoning_effort = "none" with temperature T = 0.7, consistent with (6) (9). The choice of disabling the reasoning mode reflects an API constraint: OpenAI's reasoning models do not accept the temperature parameter when reasoning_effort is set to "medium" or "high" (12). Setting reasoning_effort = "none" is therefore the only configuration in which T can be explicitly fixed to a value comparable across studies and across the Gemini models tested here. We note that enabling reasoning would likely improve GPT-5.2's performance on specific items; a sensitivity analysis at reasoning_effort = "medium" and "high" is reported in Section 5, where the trade-off between configuration parity and reasoning depth is discussed in detail.

All three models are multimodal, meaning they can process both text and images as input. This is essential because the survey items contain visual scenarios that must be interpreted to answer the questions.

## 2.3 Research questions

The following research questions guide our analysis:

**RQ1: How do frontier LLMs perform on the CRCI, a recently developed concept inventory unlikely to be present in their training data?**
The CRCI was not publicly available at the time of testing. This allows us to assess LLM performance in a setting where high accuracy cannot be attributed to memorization of items from training data.

**RQ2: When the models answer incorrectly, what types of errors emerge, and how do error profiles differ across GPT-5.2, Gemini 3 Pro, and Gemini 3 Flash?**
Following Polverini and Gregorcic (6), we classify errors observed in the chain-of-thought output as arising from: (a) incorrect interpretation of the visual content of the item (visual errors), (b) incorrect application of physics principles, evaluated independently of the correctness of the visual interpretation (reasoning errors) , or (c) inconsistencies between the reasoning and the selected answer (coordination errors). By testing three models under identical conditions, we can both characterize the nature of the difficulties and identify systematic differences across architectures. This multi-model comparison adds a dimension that is generally absent from the existing literature, which has predominantly focused on OpenAI models.

**RQ3: How does LLM performance compare to that of university students, and do the models gravitate toward the same distractors?**
We compare the models' overall scores and item-level performance to the student data reported in (11) (N = 267 first-year physics students). Beyond aggregate performance, we examine whether the incorrect answer options preferred by the LLMs correspond to the distractors most frequently selected by students. This comparison probes whether LLM errors reflect patterns similar to student misconceptions or arise from qualitatively different mechanisms. As noted in (9), care is needed in interpreting such comparisons, since the cognitive processes underlying student and LLM responses are fundamentally different — a point reinforced by the case of Q1 in the CRCI, where GPT-5.2 fails systematically due to a translation artifact[1].

# 3. Methods

## 3.1 Data collection

Each of the 21 CRCI items was submitted as a high-resolution screenshot of the English translation (11), replicating the format a student would encounter on paper. The items were submitted individually — one item per API call — to each of the three models via their respective APIs.

[1] The English translation of Q1 uses the term "comoving", which in physics carries relativistic connotations absent in the original Italian "solidale al corpo". With this wording, GPT-5.2 selected the incorrect answer in 97% of iterations, on an item that 87% of students answer correctly. To isolate this linguistic effect, we re-ran Q1 for all three models replacing"comoving" with "attached", and report only these results.

The prompt was designed to elicit a structured chain-of-thought response. Each model was instructed to: (1) describe the visual content of the image (Visual Phase), (2) generate a step-by-step derivation about the physics of the problem (Reasoning Phase), and (3) provide a final answer letter (Answer Phase). The output was returned as a structured JSON file using the API's structured output functionality, ensuring consistent parsing across all responses. This three-phase structure, consistent with the chain-of-thought approach used in (6) (9), was essential for our qualitative analysis, as it allows the source of errors to be traced to a specific stage of the response.

The temperature was set to 0.7 for all three models, consistent with the default used in (6) (9). For GPT-5.2, the reasoning mode was disabled (reasoning_effort = "none"), since the OpenAI API does not allow temperature control and reasoning to be active simultaneously (see section 2.2). Each API call was independent: no conversation history was carried over between iterations.

Google's Gemini 3 Developer Guide recommends T=1.0 as the default for the Gemini 3 family (13). To verify the robustness of our findings against this alternative configuration, we performed a full replication at T=1.0 for both Gemini 3 Flash and Gemini 3 Pro; the results are reported in Supplementary Material and indicates that the qualitative error patterns described in the following sections are unaffected by this choice

Each item was submitted 30 times per model, yielding 630 responses per model, for a grand total of 1890 responses. Data were collected between December 20, 2025 and January 10, 2026.

## 3.2 Qualitative coding of responses

All responses — not only incorrect ones — were coded qualitatively. The decision to code all responses, including those with a correct final answer, was motivated by the well-documented phenomenon of false positives in LLM outputs: cases where the model arrives at the correct answer despite flawed reasoning (6). Conversely, false negatives — correct reasoning leading to an incorrect answer — can also occur, for example when the model correctly interprets a visual scenario, but then selects the wrong letter.

The coding protocol was developed starting from the framework introduced by Polverini and Gregorcic (6) for the BEMA, and adapted to the specific characteristics of the CRCI. Each response was evaluated on three binary dimensions:

- **Visual score (V0/V1):** Does the model correctly describe the visual content of the item? This includes correctly parsing the physical scenario, the spatial arrangement of objects, the reference frames involved, and any graphical elements (trajectories, arrows, diagrams).
- **Physics reasoning score (P0/P1):** Does the model apply the relevant physics correctly? This evaluates the identification and application of physical principles (e.g. Galilean transformations, fictitious forces, the equivalence principle), independent of whether the visual interpretation was correct.

- **Coordination score (C0/C1):** Is the final answer logically consistent with the stated reasoning? This parameter assesses the internal coherence of the response — whether the selected letter is the logical consequence of what the model claimed to see and to know, regardless of whether the reasoning itself is correct.

This structured output allows us to decouple visual encoding from physical reasoning — a separation that is not possible when only the final answer is evaluated — and constitutes the methodological basis for the error analysis in section 4.

Initial calibration of the coding protocol was conducted jointly on a subset of items. After calibration, the full dataset was coded. We emphasize that our coding classifies the model's textual output, not its internal processes. A 'visual error' means the written description of the scene is incorrect, not that the model's perceptual mechanisms failed. This is standard practice in the LLM evaluation literature (6), where chain-of-thought outputs are treated as observable behavior, without claims about underlying cognition.

# 4. Results and discussion

The structured three-phase output of each response (Visual Phase, Reasoning Phase, Answer Phase) allows us to go beyond binary accuracy and distinguish the model's ability to extract features from the item's visual content, from its ability to apply physics principles. This separation — which is not possible when only the final answer is evaluated — is the basis of the analysis that follows.

## 4.1 Overall performance

Figure 1 shows the accuracy for each CRCI item — expressed as percentage of correct responses out of 30 iterations — for all three models alongside the student sample from (11).

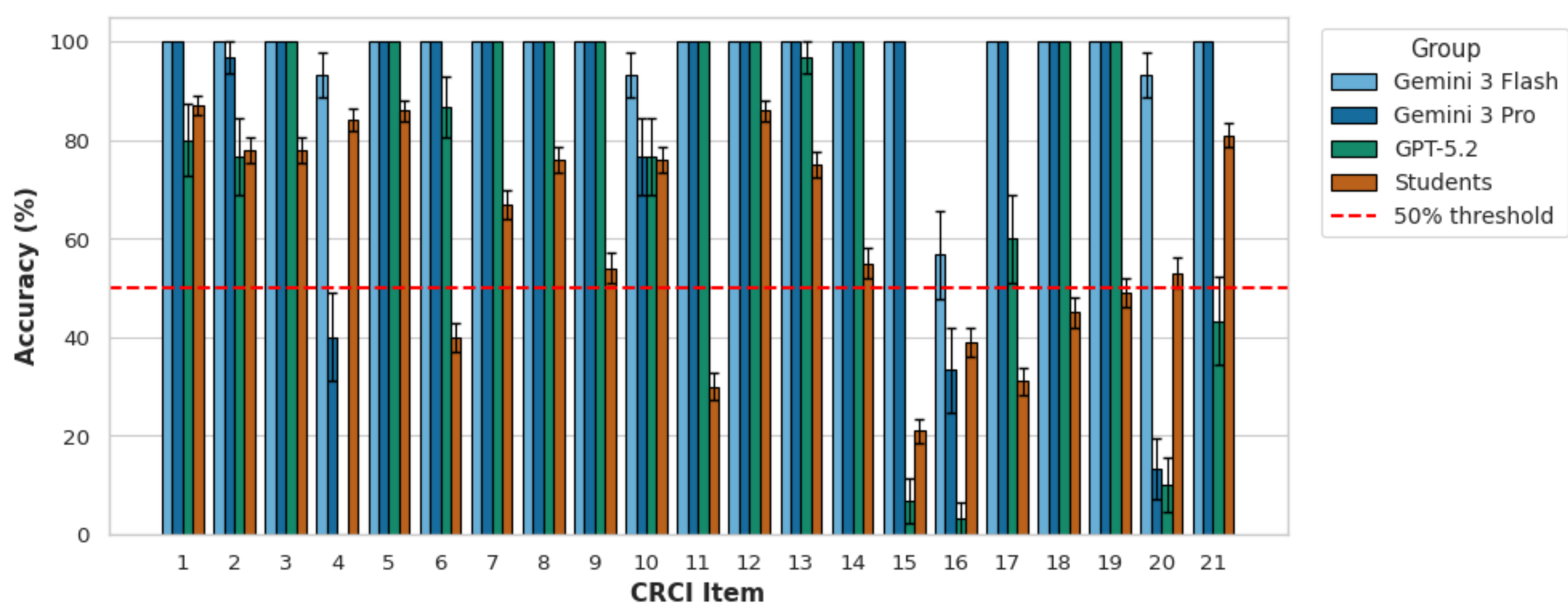

*Figure 1. Item-level accuracy on the 21 CRCI items for the three LLMs and for the student sample from (11). For each LLM, accuracy is defined as the percentage of correct responses out of 30 independent iterations at temperature T = 0.7. For GPT-5.2, the reported results refer to the baseline configuration (reasoning_effort = "none"). For students (N = 267 first-year physics students), accuracy is the percentage of correct responses. Numerical values are reported in Table S1.*

The most striking feature is that all three models achieve near-perfect accuracy on a large subset of items. Items Q3, Q5, Q7, Q8, Q9, Q11, Q12, Q13, Q14, Q18, and Q19 are answered correctly in close to 100% of iterations by all models. These items span different conceptual areas of the CRCI (inertial frames, Galilean velocity composition, free-fall scenarios, orbital motion), indicating that the models possess robust knowledge of the underlying physics when the visual component is straightforward or does not affect the response.

However, a number of items reveal clear and systematic difficulties. The pattern of failure differs across models, and three clusters can be identified:

(a) Items involving fluids on inclined planes (Q15, Q16): These items involve the position of a fluid on an inclined plane viewed from a non-inertial reference frame. Q16 (with friction) is the only item on which all three models show low accuracy (Flash 57%, Pro 33%, GPT-5.2 3%), suggesting that the visual interpretation of this scenario — which requires coordinating a 3D perspective with a frame transformation — poses a fundamental challenge for current multimodal architectures. Q15 (frictionless case), by contrast, reveals a model-specific failure: both Gemini models answer correctly in all 30 iterations, while GPT-5.2 selects distractor E in 93% of cases — a failure rate even higher than the student average of 21% on the same item. Interestingly, this failure is partially resolved when reasoning is enabled: GPT-5.2's accuracy on Q15 rises from 7% to 90% at reasoning_effort = 'high' (Table S1).

(b) Items where specific models fail (Q4, Q17, Q20, Q21*):* Q4 (Galilean velocity composition) is particularly revealing: Gemini 3 Flash answers correctly in nearly all iterations, while GPT-5.2 scores 0%, with all incorrect responses clustering on the same wrong option (see section 4.3 table 2). Q17 (pendulum in free fall) shows a different pattern: both Gemini models answer correctly in all iterations, while GPT-5.2 selects the correct answer in only 60% of cases (see section 4.3 table 2). Q20 and Q21 also show model-specific failures, with GPT-5.2 scoring 10% and 43% respectively, while Flash remains above 93% on both.

(c) The Q1 translation artifact: The English translation of Q1 originally used the term "comoving", which led GPT-5.2 to systematic errors (see footnote 1). After replacing "comoving" with "attached" — the results reported throughout this paper — GPT-5.2 improved substantially but still shows reduced accuracy (80%), suggesting that some difficulty on this item persists beyond the linguistic artifact. Both Gemini models are unaffected by the wording change, scoring 100% in both conditions.

Table S1 in the supplementary material reports Item-level accuracy (%) for each model and for the student sample.

Across the 21 items, Gemini 3 Flash slightly outperforms the larger Gemini 3 Pro, with mean accuracies of 97.0% and 88.6%, respectively (see Table 1). There is no item on which Pro performs better than Flash (see Table S1). This is consistent with an inverse-scaling pattern previously reported in the AI benchmarks literature (14). A robustness check at T=1.0, Google's recommended default for Gemini 3 (see Table S2 in the supplementary material), reduces the gap but does not eliminate it, indicating that the difference is not an artifact of the temperature setting.

When compared to students, the LLMs outperform the student average on most items. However, the exceptions are significant. On Q4, students score approximately 84% while GPT-5.2 scores 0%, due to a systematic visual error, as better discussed in section 4.3.

On Q16, both students and all three models perform poorly, but students' errors are distributed across several distractors while the models tend to concentrate on a single wrong option. On Q17, the student average is modest (~31%), and GPT-5.2 also shows difficulty (60% accuracy), while both Gemini models answer correctly in all iterations.

These cases illustrate that high overall AI accuracy can coexist with complete failure on specific items — a concern for instructors who might rely on chatbots as tutoring tools.

We note that GPT-5.2 was tested with reasoning disabled (see section 2.2). When reasoning is enabled, its mean accuracy rises to 85–86%, approaching that of Gemini 3 Pro (see Table S1 and section 5).

***Table 1.*** *Overall performance on the CRCI. Mean accuracy and standard deviation are computed across the 21 items. For each LLM, the accuracy on a given item is the fraction of correct responses out of 30 independent iterations. For students, the accuracy on each item is the percentage of correct responses from N = 267 first-year physics students (11). Note that for LLMs the SD reflects inter-item variability, whereas for students it reflects inter-subject variability; the two measures are not directly comparable. The last two columns report the number of items answered correctly in all iterations (100%) and those below the 50% threshold. For GPT-5.2, results reflect the baseline configuration (reasoning disabled). A full sensitivity analysis of GPT-5.2 with reasoning enabled is provided in Section 5 and Table S1.*

| **Model** | **Mean accuracy (%)** | **SD (%)** | **Items at 100%** | **Items at < 50%** |
|---|---|---|---|---|
| Gemini 3 Flash | 97.0 | 9.5 | 17 | 0 |
| Gemini 3 Pro | 88.6 | 25.9 | 16 | 3 |
| GPT-5.2 | 73.3 | 37.2 | 10 | 5 |
| Students (N = 267) | 61.5 | 21.3 | 0 | 7 |

## 4.2 Error profiles

Figure 2 presents the error composition for each item and each model, classified through qualitative coding into visual errors, reasoning errors, and coordination errors.

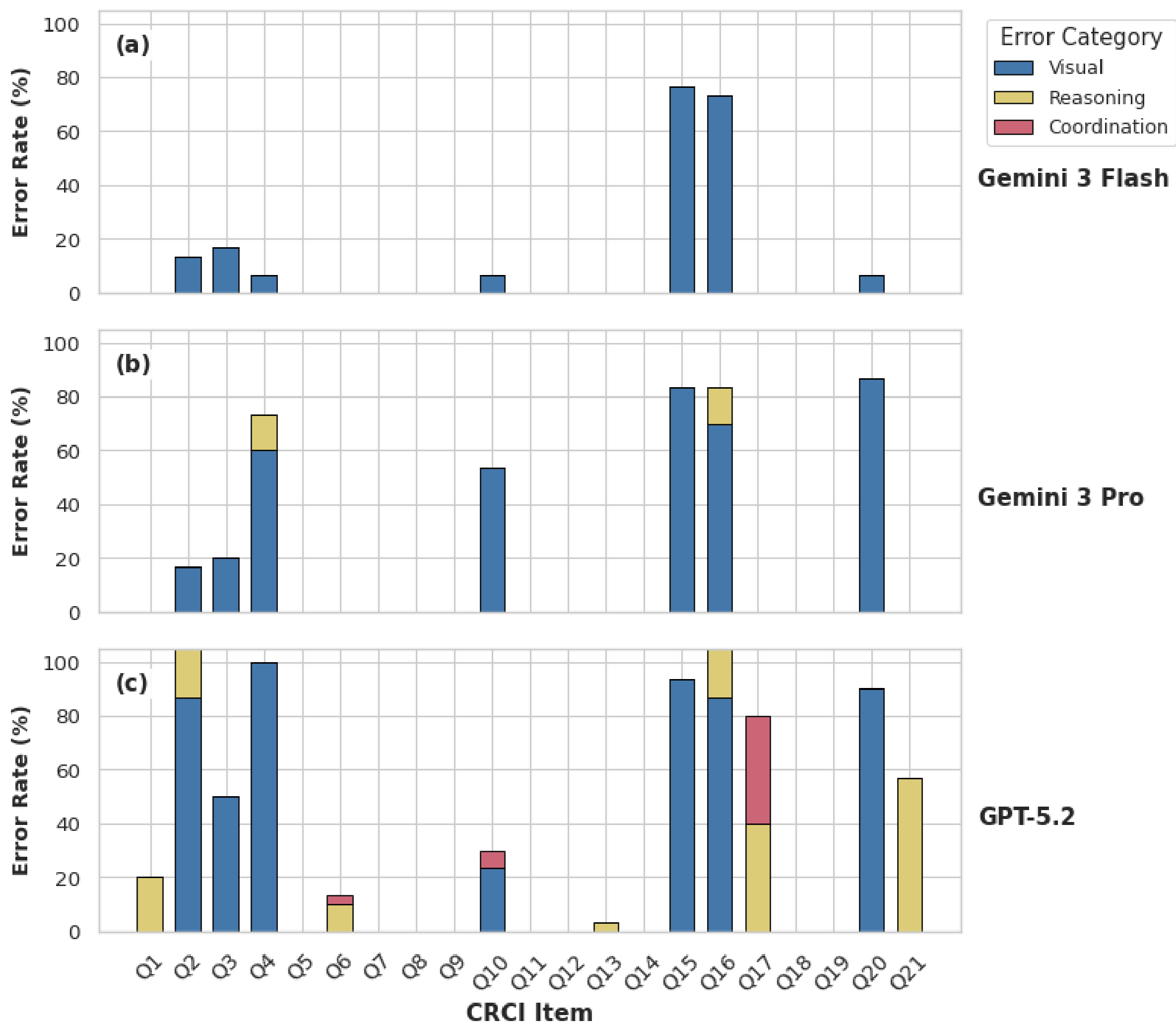

***Figure 2****. Error composition for each CRCI item, by model. Stacked bars show the breakdown of coded responses into visual errors only (V0, P1), reasoning errors only (V1, P0), combined visual and reasoning errors (V0, P0), and coordination errors (V1, P1, C0), based on the qualitative coding described in §3.2. Panels: (a) Gemini 3 Flash, (b) Gemini 3 Pro, (c) GPT-5.2 (baseline configuration, reasoning_effort = "none").*

**Gemini 3 Flash (panel a)** produces few incorrect responses which are all classified as visual errors. When Flash answers incorrectly, it does so because it misinterprets the image, not because it lacks the relevant physics knowledge. Items showing elevated visual error rates in the coding — including cases where the model reaches the correct answer despite a flawed visual description — are Q2, Q3, Q4, Q10, Q15, Q16, and Q20, all of which involve diagrams with spatial or perspective complexity. Particularly, for Q15, Flash is coded with a visual error in 23 out of 30 responses, yet it answers correctly every time because the visual error affects one of the wrong answers. Reasoning and coordination errors are essentially absent.

**Gemini 3 Pro (panel b)** shows a different pattern. Visual errors remain the dominant category, but reasoning errors appear on two items: Q4 and Q16. On Q4, 14 out of 18 incorrect responses are classified as pure visual errors (V0, P1), while the remaining 4 combine visual and reasoning errors (V0, P0). Q16 shows a similar pattern, with 16 visual errors and 4 visual-plus-reasoning errors. On Q2, Q3, Q10, and Q20, errors are exclusively visual. As with Flash, Q15 shows a high visual error rate in the coding (25 out of 30 responses coded V0) despite 100% accuracy — a compensatory pattern even more pronounced than in Flash. The overall error rate is substantially higher than for Flash, indicating the inverse scaling phenomenon [McKenzie2023] at the qualitative level.

**GPT-5.2 (panel c)** presents the most complex error profile, with errors across all three categories. Reasoning errors dominate on Q1, where a form of semantic misalignment persists even with the corrected wording (see footnote 1), and on Q21, where all 17 errors are coded (V1, P0). In Q2, visual errors are dominant, with 22 out of 29 coded errors classified as V0. Seven responses involve reasoning errors, including 4 coded (V0, P0) and 3 coded (V1, P0). Within the (V0, P0) category, Q2 also includes two false-positive cases, in which the chatbot selects the correct final answer despite producing reasoning that is not fully coherent with that answer. The two false positives are coded (V0, P1, C0) and (V0, P0, C0). Although they have different codes, both cases appear to stem from the same underlying pattern: the chatbot initially fails to interpret the schematic arrangement of the diagram correctly, but then redescribes it correctly at the end of the physics reasoning phase, leading it to select the right final answer. On Q4, GPT-5.2 fails entirely, but the qualitative coding reveals that all 30 errors are just visual (V0, P1, C1): the model misreads the spatial arrangement of the scenario and then applies correct physics to its flawed geometric premise. Visual errors also dominate on Q15, Q16, and Q20. In particular, on Q16 a false positive occurs, in which flawed reasoning does not affect the selection of the correct answer. Q17 stands apart: here, GPT-5.2's 12 errors are classified as combined reasoning and coordination failures (V1, P0, C0), meaning the model correctly interprets the visual scenario but applies incorrect physics and then selects an answer inconsistent even with its own flawed reasoning. Smaller coordination errors also appear on Q6 (1 out of 4 errors) and Q10 (2 out of 7 errors).

From this analysis, four main failure modes can be identified:

- **Semantic misalignment:** The model's response is driven by a linguistic interpretation that diverges from the intended physical meaning. The clearest example is Q1, where GPT-5.2 interprets the terms in a specialized terminological sense found in certain treatments of special relativity (15).
- **3D spatial vision failure:** The model fails to correctly parse diagrams involving perspective, inclined planes, or reference frame transformations represented visually. This is the dominant failure mode on Q16 across all models, and on Q15 for GPT-5.2.
- **False positives:** The model arrives at the correct answer despite flawed visual interpretation or incorrect reasoning. These were detected by coding all responses, not only incorrect ones.
- **Coordination failure:** The model describes the scenario correctly, generates an adequate physics derivation, but selects a letter that does not match its own conclusion. This mode is rare but was observed in GPT-5.2, particularly on Q17.

The overall picture supports the conclusion that, on the CRCI, multimodal failures stem primarily from visual misinterpretations rather than from deficits in physics knowledge. This is most clearly demonstrated by the Flash model, whose errors are almost entirely visual, but the pattern holds for the other models as well: reasoning errors, when they occur, are concentrated on a small number of items and are often entangled with visual difficulties.

## 4.3 Distractor analysis and comparison with students

Figures S1 and S2 in supplementary material show the distribution of selected answer options for each CRCI item, comparing the three models with the student population from (11). The correct answer is marked with an asterisk. In table 2 below it is shown the most selected distractor for each question and each model and students.

***Table 2.*** *Most frequently selected distractor for each CRCI item, by model and by students. For each LLM, the letter indicates the most selected incorrect option across 30 iterations, with the percentage of all responses in parentheses. For students (N = 267), the most selected incorrect option is reported with the percentage of students who chose it (11). A dash (—) indicates that all responses were correct. The correct answer for each item is listed in the Key column. The full response distributions are available in the supplementary material.*

| Item | Key | Gemini 3 Flash | Gemini 3 Pro | GPT-5.2 | Students |
|---|---|---|---|---|---|
| Q1 | A | — | — | B (20%) | B (13%) |
| Q2 | B | — | A (3%) | D (23%) | A (15%) |
| Q3 | C | — | — | — | A (18%) |
| Q4 | A | C (7%) | C (40%) | C (100%) | B (9%) |
| Q5 | B | — | — | — | A (14%) |
| Q6 | C | — | — | B (13%) | B (43%) |
| Q7 | B | — | — | — | C (26%) |
| Q8 | A | — | — | — | B (19%) |
| Q9 | C | — | — | — | A (38%) |
| Q10 | B | A (7%) | A (23%) | A (17%) | A (12%) |
| Q11 | A | — | — | — | C (40%) |
| Q12 | B | — | — | — | A (10%) |
| Q13 | D | — | — | B (3%) | B (24%) |
| Q14 | C | — | — | — | B (25%) |
| Q15 | B | — | — | E (93%) | A (30%) |
| Q16 | E | D (27%) | D (43%) | A (70%) | C (24%) |
| Q17 | C | — | — | B (40%) | A (27%) |
| Q18 | E | — | — | — | A (32%) |
| Q19 | C | — | — | — | B (19%) |
| Q20 | C | B (3%) | B (87%) | B (47%) | A (42%) |
| Q21 | C/E | — | — | D (57%) | A (10%) |

On items where all models perform well (e.g., Q5, Q7, Q8, Q9, Q11, Q12, Q13), the distractor distributions are highly concentrated on the correct option: models select the correct option in nearly all iterations, and students show a broader but still predominantly correct distribution.
We note that in Q3, models correctly select C as their final answers (see Table S1), even though all three LLMs exhibit a visual error (Figure 2), because the misinterpretation of the spatial arrangement does not affect the answer to this question.

Q4. Roberto is sitting on a bridge and sees Alice passing by on a boat traveling at a constant speed along the river. Just as Alice is passing by, a drone D starts from point P and flies at a constant speed to point Q on the opposite bank. How intense will the drone's speed be in Alice's frame of reference?

(A) Greater than the magnitude of the drone's velocity in Roberto's frame of reference.

(B) Equal to the magnitude of the drone's velocity from Roberto's frame of reference.

(C) Impossible to determine without knowing the magnitude of the boat's and drone's velocities.

(D) Less than the magnitude of the drone's velocity from Roberto's frame of reference.

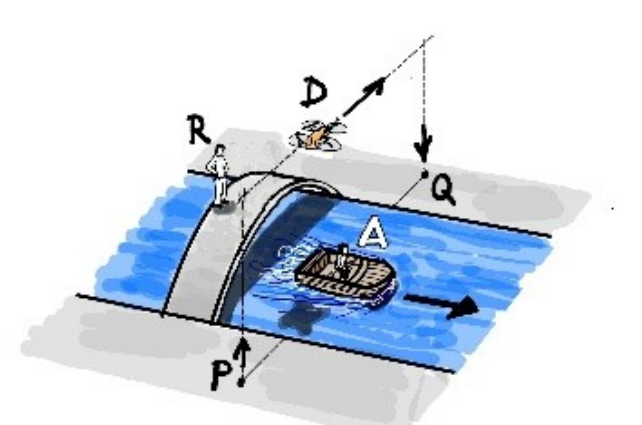

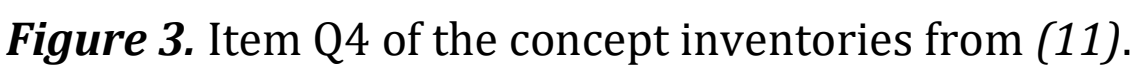

***Figure 3.*** Item Q4 of the concept inventories from *(11)*.

The most informative cases are those where errors occur.

We underscore five items:

**Q4 (Galilean velocity composition):** This is the most impressive divergence between models. The item tests conceptual understanding of Galilean relative velocity by requiring a vector — rather than scalar — interpretation of motion and a clear geometric recognition of perpendicular directions: subtracting the boat's velocity from the drone's velocity produces a resultant whose magnitude necessarily increases, since the vector difference introduces an additional orthogonal component (Figure 3). The result can therefore be inferred qualitatively, without any numerical calculation.
As reported in Table S1, Gemini 3 Flash selects the correct answer (A) in 93% of iterations, closely matching students' performance (about 85%). By contrast, GPT-5.2selects option C in 100% of iterations — a complete failure with zero variance — and Gemini 3 Pro is split across A (40%), C (40%), and D (20%). In particular, the distractor C corresponds to the claim that the drone's speed in Alice's frame cannot be determined without knowing the numerical values of the velocities. The correct reasoning requires recognizing that the drone flies perpendicularly to the boat's direction, so the Galilean velocity sum always has a greater magnitude than the drone's velocity alone — regardless of the specific values. GPT-5.2's behaviour here resembles a systematic error rather than stochastic error. In fact, qualitative analysis (section 4.2) reveals it stems from a visual error: the model misreads the spatial arrangement in the diagram, interpreting the drone's trajectory as having a component along the boat's direction, and then applies the Galilean transformation correctly to its faulty geometric premise. Gemini 3 Pro's selection of D stems from the same visual error; in two-thirds of those cases, however the model also introduces a hallucinated assumption that, in physics problems, boats are generally slower than a flying drone.

**Q15 (fluid on frictionless inclined plane):** shows a striking divergence: both Gemini models answer correctly in all iterations, while GPT-5.2 selects distractor E in 93% of cases.

**Q16 (fluid on inclined plane with friction)** is the only item where all three models struggle, suggesting that this specific visual scenario poses a fundamental challenge for current architectures.

**Q17 (pendulum in free fall):** Both Gemini models select the correct answer (C, uniform circular motion) in all iterations. GPT-5.2 selects C in 60% of cases, correctly recognizing that the rigid rod provides centripetal acceleration in the zero-gravity frame. In the remaining 40%, the same model dismisses this possibility — arguing that centripetal acceleration "cannot be sustained" — and concludes the trajectory is a straight line, yet selects "parabolic motion" (B) as "the closest description". This 60/40 split on identical input reveals that GPT-5.2's grasp of mechanical constraints under effective weightlessness is fragile rather than absent.

**Q21 (orbital motion):** This item accepts two correct answers (C and E). Flash gravitates strongly toward C, Pro toward E, and GPT-5.2 distributes across C, D, and E. The student distribution is broadly similar, with the majority on C and E. Notably, GPT-5.2 places a substantial fraction on D, a distractor that students rarely select, suggesting a model-specific error pattern unrelated to student reasoning.

Overall, the distractor analysis indicates that LLM error profiles do not reproduce students' common difficulties patterns. When models err, they tend to concentrate on a single distractor with high consistency (low variance), whereas students distribute their errors more broadly. Moreover, the specific distractor preferred by a failing model is often different from the one most frequently selected by students. This finding is consistent with the observation that model errors are predominantly visual or semantic in origin, rather than reflecting the conceptual difficulties that drive student misconceptions; this is also in line with (9). Item-by-item distributional comparisons remain qualitative and descriptive, consistent with the approach adopted in (6): with 30 iterations per model per item, the per-item distributions do not support robust hypothesis testing. An aggregate test across items is however feasible: using Shannon entropy (16) as a per-item dispersion measure, a paired Wilcoxon signed-rank test across the 21 items confirms that LLM response distributions are systematically more concentrated than student responses ($p < 0.005$ for all three models).

## 4.4 Implications for teaching

The results presented above have direct implications for the use of AI chatbots in physics education.

First, the high overall accuracy of all three models on the majority of CRCI items may give students — and instructors — a misleading impression of reliability. A chatbot that answers 18 out of 21 questions correctly appears competent, but the remaining three failures can be severe and unpredictable. On Q4, for instance, GPT-5.2 provides a confidently wrong answer in every single iteration. A student relying on this tool for studying Galilean velocity composition would consistently receive incorrect guidance, delivered with the same apparent confidence as a correct response. Instructors should be aware of this pattern and communicate it to students: chatbot accuracy is item-dependent, and high average performance does not guarantee reliability on any specific question.

Second, the prevalence of visual errors has pedagogical consequences. Many of the CRCI items that challenge the models involve hand-drawn illustrations, spatial perspectives, and reference frame diagrams — precisely the kind of representations that are central to physics instruction. If chatbots struggle with these, their usefulness as tutors on topics that require interpreting experimental setups, free-body diagrams, or frame transformations is limited (6). This weakness is not something that students can easily detect: the model's textual reasoning may appear sound even when its visual interpretation is wrong.

Third, the existence of false positives — correct answers produced by flawed reasoning — represents a subtle but important risk. A student who asks a chatbot to explain its reasoning on a question it answered correctly may receive a plausible but physically incorrect explanation. This could reinforce misconceptions rather than address them. An example can be observed in the response to item Q16 provided by ChatGPT 5.2, which asks for the inclination of the water surface inside a container sliding down a rough inclined plane. According to the chatbot's reasoning, the water surface should exhibit an inclination even steeper than that of the plane itself—a physical impossibility, given that only gravity and kinetic friction are acting on the system. Nevertheless, in its final response, the model identifies the correct alternative due to a visual error.

Fourth, the difference in performance between the lighter and the more expensive model, where Gemini 3 Flash performs slightly better than the larger Gemini 3 Pro across both temperature configurations tested, suggests that instructors and students should not assume that a more advanced or more expensive model is necessarily more reliable for physics tasks. This observation is consistent with the cost–performance analysis reported by Polverini and Gregorcic (7), who found that higher-priced models do not always yield better results on physics concept inventories. Model selection should be guided by empirical evaluation on the specific domain, not by marketing claims or general benchmarks.

Finally, the high accuracy demonstrated by LLMs on the majority of CRCI items has implications for how concept inventories are administered. If chatbots can achieve near-perfect scores on most items, take-home administration of such instruments is no longer a reliable measure of student understanding. Instructors who use concept inventories for formative or summative assessment should administer them in supervised, in-person settings to ensure that the responses reflect genuine student reasoning.

These findings call for critical awareness of chatbot limitations, not for avoiding them entirely, particularly on tasks involving visual interpretation and conceptual reasoning about reference frames and relative motion.

# 5. Limitations and Robustness

Several limitations of this study should be acknowledged.

First, our results represent a snapshot of the current state of AI technology. LLMs are evolving rapidly, and future models may overcome the visual and reasoning difficulties documented here. This study should therefore be read as a historical reference point, against which future progress can be measured — much as earlier evaluations of ChatGPT-3.5 on the FCI now serve as benchmarks for how quickly the technology has advanced (6).

Second, several configuration choices warrant attention.
The temperature parameter was set to T=0.7 across all models, while Google's recommended default for the Gemini 3 family is T=1.0. A sensitivity analysis at this alternative setting is reported in Table S2 of the supplementary material, and confirms that the qualitative findings of this study are not affected.
Furthermore, the three models were not tested under strictly equivalent configurations. All models were queried at T=0.7, but for GPT-5.2 this required disabling the reasoning mode (reasoning_effort = "none"), as the OpenAI API does not allow temperature control and reasoning to be active simultaneously (12).

The Gemini 3 models were queried without specifying a thinking level, which defaults to "high" in the Google API [ref-Google-devblog-2025]. This means that the Gemini models operated with their full reasoning capabilities enabled, while GPT-5.2 did not — an asymmetry that reflects each provider's API design rather than a methodological oversight. We note that this configuration corresponds to the default operational profile that a typical user would encounter when interacting with each model, which is the scenario most relevant to the pedagogical concerns motivating this study. Nevertheless, cross-model accuracy comparisons should be interpreted with this asymmetry in mind; our primary focus is on characterizing error types rather than ranking models.

A related limitation concerns the rapid evolution of the LLM landscape. The original gemini-3-pro-preview model used in our main data collection (released 18 November 2025) was deprecated by Google on 9 March 2026, less than four months later (17). The robustness check at T=1.0 reported in Table S2 therefore uses the successor model gemini-3.1-pro-preview (released 19 February 2026), with thinking_level set to "medium", the closest available equivalent to the original model configuration. While the Gemini 3 Flash model used in this study remains available, the deprecation of the Pro model after less than four months of availability illustrates a broader methodological concern for PER-on-LLM research: studies of this kind are inevitably time-stamped, and exact replication is sometimes materially impossible because the systems being characterised cease to exist. To enable post-hoc verification, the raw model outputs from this study will be made available in a public repository upon publication (18).

To address the configuration asymmetry previously described, we assessed the sensitivity of GPT-5.2's performance to reasoning depth by repeating the 30-iteration protocol at reasoning_effort = "medium" and "high" (Table S1). At these settings the temperature parameter is not user-configurable (12). Items whose baseline errors were classified as physics failures improved substantially at "medium" and held at "high" (e.g., Q17: 60% → 100% → 100%; Q21: 43% → 97% → 93%). Items dominated by visual errors showed a more complex pattern: Q15 improved monotonically (7% → 60% → 90%), while Q4 remained near zero across all settings, confirming a visual bottleneck that reasoning cannot overcome. On several items (Q16, Q20), accuracy peaked at "medium" and decreased at "high", indicating that excessive reasoning can degrade performance on visually demanding items. Overall, the sensitivity analysis reveals a clear interaction between error type and reasoning depth: items dominated by physics errors are resolved by moderate reasoning, items with pure visual errors are unaffected regardless of reasoning level, and items with complex visual demands show non-monotonic behavior where excessive reasoning can degrade performance.

Indeed, when reasoning is enabled, GPT-5.2's mean accuracy (85–86%) approaches that of Gemini 3 Pro (89%) (see table S1), suggesting that much of the apparent performance gap in the baseline comparison reflects the configuration asymmetry rather than a fundamental difference in physics capabilities.

Third, we tested a single concept inventory covering a specific area of physics (classical relativity). While the CRCI spans a range of conceptual areas — reference frames, Galilean transformations, and the equivalence principle — our findings may not generalize to other topics or to other types of assessment instruments. In particular, the strong role of visual errors in our results is likely related to the specific nature of the CRCI diagrams, which involve 3D perspectives, inclined planes, and reference frame transformations. Instruments with different visual demands may produce different error profiles.

Fourth, the qualitative coding of responses, while grounded in the framework of Polverini and Gregorcic (6), inevitably involves a degree of subjectivity. The binary scoring (0/1) on each dimension simplifies a continuous spectrum of quality, and borderline cases require judgement calls.

Fifth, we did not explore the effect of prompt engineering on model performance. Our prompt was standardized across all three models and designed to elicit a chain-of-thought response, but more elaborate prompting strategies — such as instructing the model to describe the image in greater detail before reasoning, or providing worked examples — might improve performance on specific items. This remains an avenue for future work.

Finally, while the CRCI items were almost certainly absent from the training data, the underlying physics is standard textbook material. LLMs do not retrieve stored answers but generate responses from learned patterns, so familiarity with analogous problems may contribute to the high accuracy observed on most items. What the CRCI tests is the models' ability to handle novel formulations — new scenarios, diagrams, and distractor structures — of familiar physics. This is a qualitatively different situation from established instruments such as the FCI, whose specific items and answer keys have been widely reproduced online for decades.

# 6. Conclusions

This study addressed three research questions concerning the performance (RQ1), error profiles (RQ2), and comparison with student data (RQ3) of frontier LLMs on the CRCI. We evaluated three frontier large multimodal models, GPT-5.2, Gemini 3 Pro, and Gemini 3 Flash, on the Classical Relativity Concept Inventory (CRCI), a recently developed and validated instrument that was not publicly available at the time of testing. By collecting 30 independent iterations per item per model and qualitatively coding all responses on three dimensions (visual interpretation, physics reasoning, and coordination), we obtained a detailed picture of how current AI chatbots handle conceptual physics tasks on an uncontaminated instrument.

All three models achieved high accuracy on the majority of CRCI items, demonstrating robust physics knowledge on topics such as inertial reference frames, Galilean velocity composition, and free-fall scenarios. However, specific items, particularly those involving 3D spatial interpretation of diagrams (Q16 for all models, Q15 for GPT-5.2) or requiring careful semantic parsing (Q1), revealed systematic and often complete failures. The qualitative analysis showed that these failures stem predominantly from visual misinterpretations rather than from deficits in physics knowledge: when the models successfully extract the visual features, they almost always generate a correct physical derivation.

Across the 21 items, Gemini 3 Flash slightly outperformed the larger Gemini 3 Pro on most items. A robustness check at the alternative temperature setting recommended by Google for the Gemini 3 family (T=1.0, see Table S2 in the supplementary material) reduces but does not eliminate this gap, indicating that the difference reflects a model-level rather than a configuration effect. This observation is relevant for instructors and students making choices about which tools to use, as it suggests that a more advanced or expensive model is not necessarily more reliable for physics tasks.

A sensitivity analysis on GPT-5.2 at three reasoning_effort levels ("none", "medium", "high") confirmed quantitatively what the qualitative error analysis showed for all three models: enabling reasoning substantially improves performance on items dominated by physics errors, but has little effect on items dominated by visual errors. Visual misinterpretation, rather than insufficient reasoning, is the primary bottleneck for current multimodal models on the CRCI.

The comparison with student data (N = 267) revealed that LLM error profiles do not reproduce the patterns of student misconceptions. When models err, they tend to converge on a single distractor with high consistency, whereas students distribute their errors more broadly across multiple options. This indicates that AI errors and student misconceptions arise from qualitatively different mechanisms. This difference is not a limitation but a crucial point: students' reasoning processes are typically heterogeneous, context-dependent, and often internally inconsistent, and it is precisely this variability that makes their thinking diagnostically accessible and pedagogically relevant. By contrast, the higher internal coherence of AI errors reflects optimization processes that do not necessarily map onto students' cognitive structures. This point should inform both the design of assessments and the pedagogical use of chatbots.

The divergence between AI errors and student errors should be regarded as a noticeable feature rather than a gap to be closed. It emphasizes that learning relies on processes — uncertainty, partial reasoning, revision, errors and even inconsistency — that are intrinsic to human cognition. This distinction should inform both the design of assessments and the cautious, well-structured integration of AI tools in educational contexts.

For physics instructors, the practical message is clear: current chatbots can be impressively accurate on conceptual physics tasks, but their reliability is item-dependent and unpredictable. High average accuracy can mask complete failures on specific questions, and correct answers may be accompanied by flawed reasoning. These tools should be used with critical awareness, and concept inventories should be administered in supervised settings rather than as take-home assignments.

Given the rapid speed of LLM development, follow-up evaluations on the same instrument will be needed to track progress. This study and the use of a novel instrument provide a baseline for tracking future AI progress on conceptual physics tasks.

# Acknowledgments

The PhD program attended by the author C.G. is financed by the European Union - Next Generation EU, Mission 4 Component 2 CUP E66E24000030008.

# Supplementary Material

This supplementary material provides the complete item-level data underlying the analysis presented in the main text. It includes the accuracy of each tested model and the student sample across the 21 CRCI items, the full distribution of selected answer options for each item, and a robustness analysis examining the effect of changing the API temperature.

## Item-Level Performance and Response Distribution

Table S1 reports the accuracy of each model and of the student sample on each of the 21 CRCI items.
Figures S1 and S2 show the full distribution of selected answer options for each item, comparing the three models with the student population from [11]. In each panel, the correct answer is marked with an asterisk (*) on the horizontal axis. For LLMs, the frequency represents the percentage of the 30 independent iterations in which a given option was selected. For students (N = 267), it represents the percentage of students who selected that option.
Table S2 reports the results of the temperature sensitivity analysis discussed in sections 3 and 5.

| | Gemini 3 Flash | Gemini 3 Pro | GPT-5.2* | GPT-5.2 (med) | GPT-5.2 (high) | Students |
|---|---|---|---|---|---|---|
| 1 | 100.0 | 100.0 | 80.0 | 96.7 | 93.3 | 87.0 |
| 2 | 100.0 | 96.7 | 76.7 | 96.7 | 96.7 | 78.0 |
| 3 | 100.0 | 100.0 | 100.0 | 100.0 | 100.0 | 78.0 |
| 4 | 93.3 | 40.0 | 0.0 | 6.7 | 13.3 | 84.0 |
| 5 | 100.0 | 100.0 | 100.0 | 100.0 | 100.0 | 86.0 |
| 6 | 100.0 | 100.0 | 86.7 | 80.0 | 90.0 | 40.0 |
| 7 | 100.0 | 100.0 | 100.0 | 100.0 | 100.0 | 67.0 |
| 8 | 100.0 | 100.0 | 100.0 | 80.0 | 93.3 | 76.0 |
| 9 | 100.0 | 100.0 | 100.0 | 100.0 | 100.0 | 54.0 |
| 10 | 93.3 | 76.7 | 76.7 | 80.0 | 70.0 | 76.0 |
| 11 | 100.0 | 100.0 | 100.0 | 100.0 | 100.0 | 30.0 |
| 12 | 100.0 | 100.0 | 100.0 | 100.0 | 100.0 | 86.0 |
| 13 | 100.0 | 100.0 | 96.7 | 100.0 | 100.0 | 75.0 |
| 14 | 100.0 | 100.0 | 100.0 | 100.0 | 100.0 | 55.0 |
| 15 | 100.0 | 100.0 | 6.7 | 60.0 | 90.0 | 21.0 |
| 16 | 56.7 | 33.3 | 3.3 | 23.3 | 13.3 | 39.0 |
| 17 | 100.0 | 100.0 | 60.0 | 100.0 | 100.0 | 31.0 |
| 18 | 100.0 | 100.0 | 100.0 | 100.0 | 100.0 | 45.0 |
| 19 | 100.0 | 100.0 | 100.0 | 100.0 | 100.0 | 49.0 |
| 20 | 93.3 | 13.3 | 10.0 | 63.3 | 50.0 | 53.0 |
| 21 | 100.0 | 100.0 | 43.3 | 96.7 | 93.3 | 81.0 |
| Mean | 97.0 | 88.6 | 73.3 | 84.9 | 85.9 | 61.5 |

***Table S1.*** *Item-level accuracy (%) for each model and for the student sample. For each LLM, accuracy is the percentage of correct responses out of 30 independent iterations. For students, accuracy is the percentage of correct responses from N = 267 first-year physics students [11]. GPT-5.2* denotes the baseline configuration (reasoning_effort = "none"), used for the qualitative error analysis in sections 4.2 and 4.3. The columns (med) and (high) report GPT-5.2 accuracy at reasoning_effort = "medium" and "high" respectively (see section 5).*

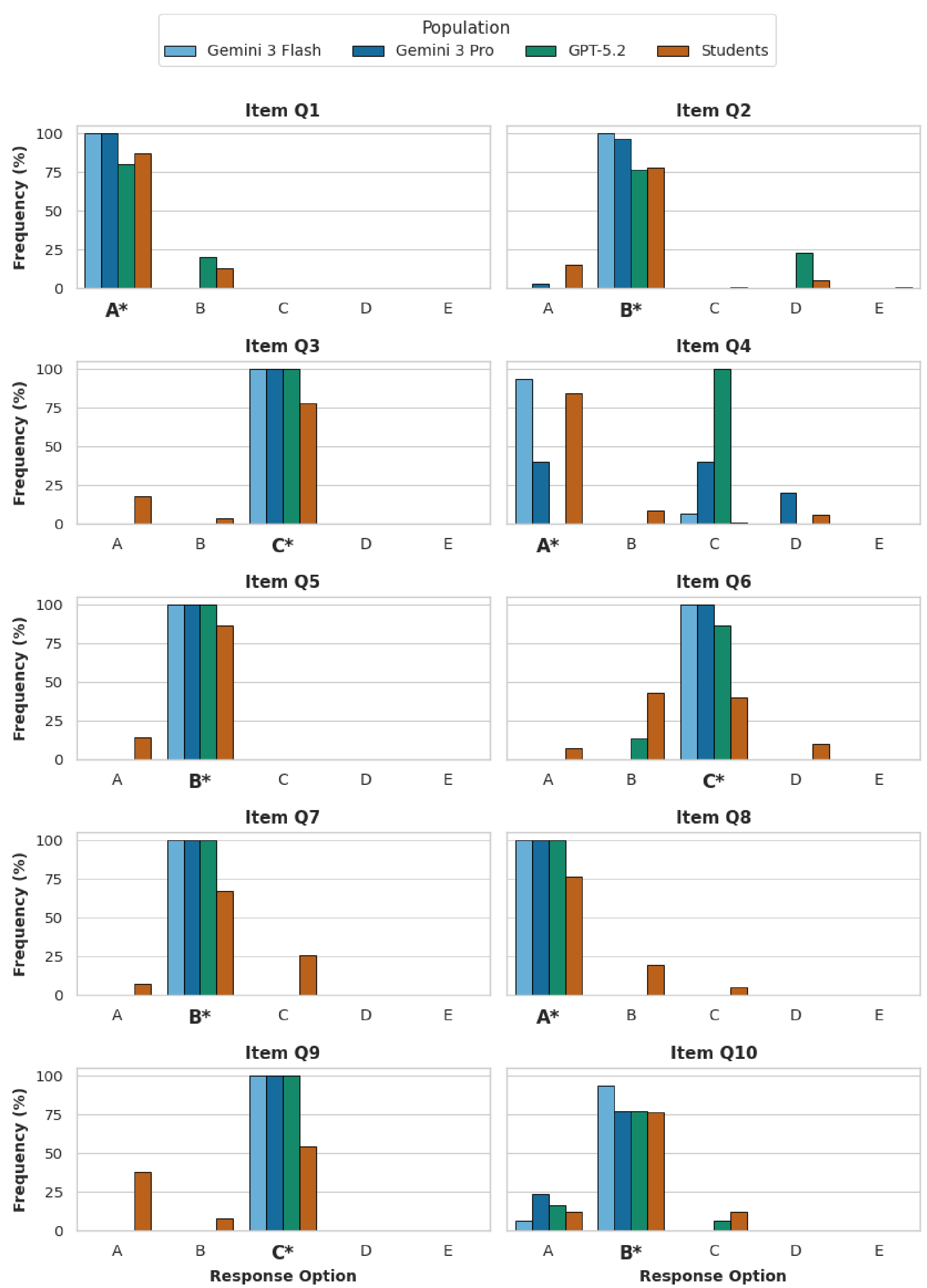

*Figure S1. Distribution of selected answer options for CRCI items Q1–Q10, for the three LLMs and for the student sample (N = 267) from [11].* An asterisk (*) indicates the correct answer. *For each model, frequencies are computed over 30 independent iterations.*

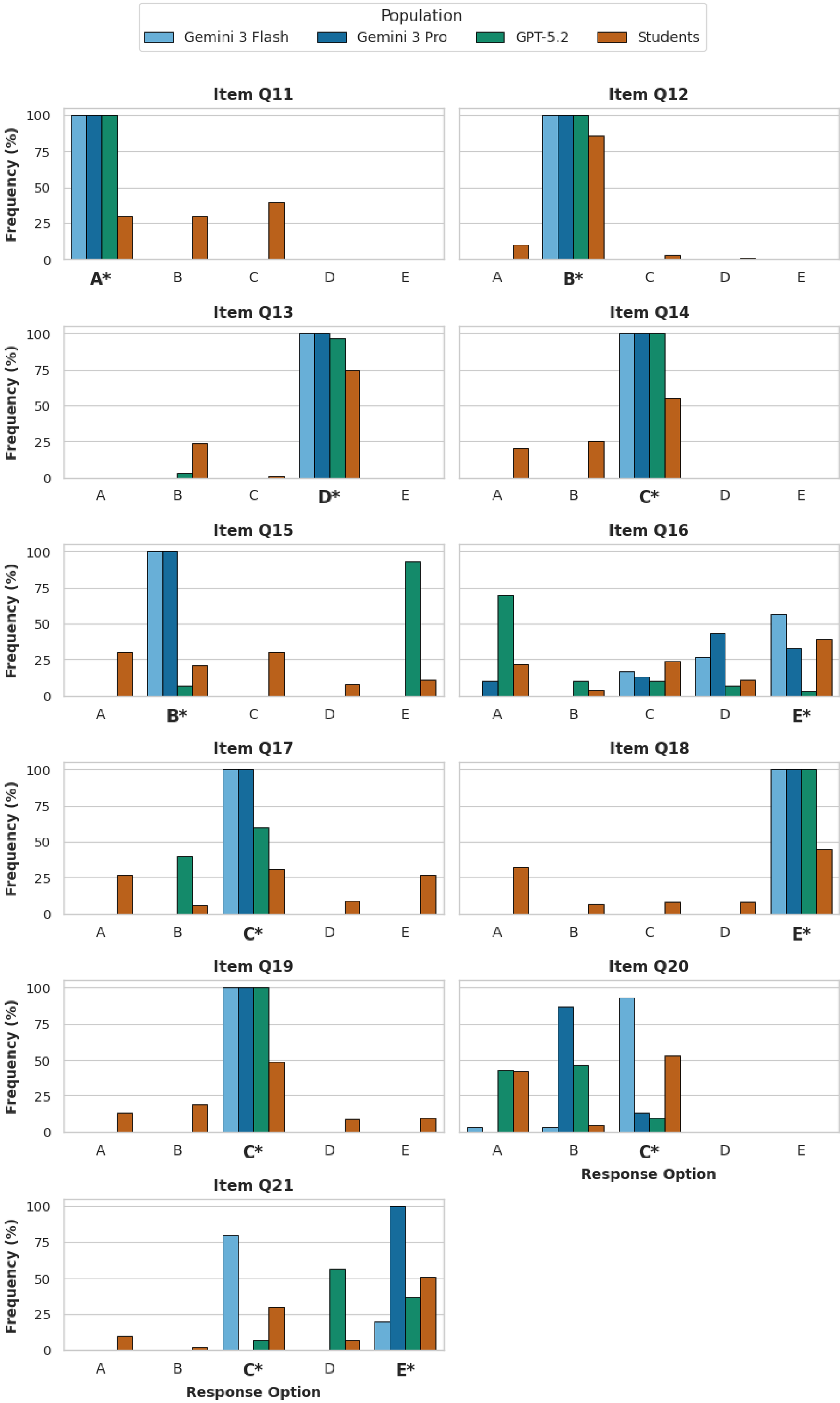
Population
Gemini 3 Flash
Gemini 3 Pro
GPT-5.2
Students
Item Q11
Item Q12
Item Q13
Item Q14
Item Q15
Item Q16
Item Q17
Item Q18
Item Q19
Item Q20
Item Q21
Frequency (%)
Response Option
A*
B*
C*
D*
E*
A
B
C
D
E
0
25
50
75
100

*Figure S2. Distribution of selected answer options for CRCI items Q11–Q21. Same format as Figure S1. Items Q15, Q16, Q17, Q20, and Q21 show the most pronounced divergence between models and between models and students (see section 4.3).*

## Robustness to API temperature

The temperature parameter T was set to 0.7 for the main study, consistent with the protocol established in (6, 9). Google's Gemini 3 Developer Guide (13) subsequently recommends T=1.0 as default for the Gemini 3 family, indicating that lower values may cause performance degradation on complex tasks. To verify the robustness of our findings against this alternative configuration, we replicated the data collection at T=1.0 for both Gemini 3 Flash and Gemini 3 Pro, with N=30 iterations per item.

The Flash replication used the same model as in the main study (gemini-3-flash-preview), varying only T as default. The Pro replication used the successor model gemini-3.1-pro-preview with thinking_level set to "medium" — the closest available equivalent to the original model configuration — because the original gemini-3-pro-preview was deprecated by Google on 9 March 2026 (17) (see section 5).

Table S2 reports the four configurations item by item. The Flash–Pro performance gap reduces but persists at T=1.0 and the difficulty of items Q4, Q16 and Q21 — are unaffected by the temperature change

| Item | Key | Flash T=0.7 | Flash T=def | Δ Flash | Pro T=0.7 | Pro 3.1 T=def | Δ Pro |
|---|---|---|---|---|---|---|---|
| Q1 | A | 100.0% | 100.0% | +0.0pp | 100.0% | 100.0% | +0.0pp |
| Q2 | B | 100.0% | 100.0% | +0.0pp | 96.7% | 86.7% | −10.0pp |
| Q3 | C | 100.0% | 100.0% | +0.0pp | 100.0% | 100.0% | +0.0pp |
| Q4 | A | 93.3% | 100.0% | +6.7pp | 40.0% | 16.7% | −23.3pp |
| Q5 | B | 100.0% | 100.0% | +0.0pp | 100.0% | 100.0% | +0.0pp |
| Q6 | C | 100.0% | 100.0% | +0.0pp | 100.0% | 100.0% | +0.0pp |
| Q7 | B | 100.0% | 100.0% | +0.0pp | 100.0% | 100.0% | +0.0pp |
| Q8 | A | 100.0% | 100.0% | +0.0pp | 100.0% | 100.0% | +0.0pp |
| Q9 | C | 100.0% | 100.0% | +0.0pp | 100.0% | 100.0% | +0.0pp |
| Q10 | B | 93.3% | 96.7% | +3.3pp | 76.7% | 100.0% | +23.3pp |
| Q11 | A | 100.0% | 100.0% | +0.0pp | 100.0% | 100.0% | +0.0pp |
| Q12 | B | 100.0% | 100.0% | +0.0pp | 100.0% | 100.0% | +0.0pp |
| Q13 | D | 100.0% | 100.0% | +0.0pp | 100.0% | 100.0% | +0.0pp |
| Q14 | C | 100.0% | 100.0% | +0.0pp | 100.0% | 100.0% | +0.0pp |
| Q15 | B | 100.0% | 96.7% | −3.3pp | 100.0% | 100.0% | +0.0pp |
| Q16 | E | 56.7% | 23.3% | −33.3pp | 33.3% | 53.3% | +20.0pp |
| Q17 | C | 100.0% | 100.0% | +0.0pp | 100.0% | 100.0% | +0.0pp |
| Q18 | E | 100.0% | 100.0% | +0.0pp | 100.0% | 100.0% | +0.0pp |
| Q19 | C | 100.0% | 100.0% | +0.0pp | 100.0% | 100.0% | +0.0pp |
| Q20 | C | 93.3% | 100.0% | +6.7pp | 13.3% | 100.0% | +86.7pp |
| Q21 | C | 80.0% | 40.0% | −40.0pp | 0.0% | 0.0% | +0.0pp |
| Avg | | 96.0% | 93.2% | −2.9pp | 83.8% | 88.4% | +4.6pp |

***Table S2.*** *Item-level accuracy (%) for Gemini 3 Flash and Gemini 3 Pro under two temperature configurations: T=0.7 (main study) and T=1.0 (Google's recommended default for Gemini 3). The Pro T=1.0 column refers to the successor model gemini-3.1-pro-preview with thinking_level = "medium". For each model–configuration combination, accuracy is the percentage of correct responses out of 30 independent iterations. Δ columns report the change in accuracy between T=0.7 and T=1.0 in percentage points.*